\documentclass[]{aa}
\usepackage{graphicx}
\usepackage{txfonts}
\usepackage{multirow}
\usepackage{gensymb}
\usepackage[authoryear]{natbib}

\usepackage{color}

\begin{document}

\title{Multi-stage four-quadrant phase mask: achromatic coronagraph\\ for space-based and ground-based telescopes}

\author{Rapha\"{e}l~Galicher\inst{1}\fnmsep\inst{2}, Pierre~Baudoz\inst{2}, Jacques~Baudrand\inst{2}}
\institute{Luth, CNRS, Observatoire de Paris, 5, place Jules~Janssen, 92195 Meudon, France\\
  \email{raphael.galicher@obspm.fr}
  \and
  Lesia, Observatoire de Paris, CNRS and University Denis Diderot Paris 7. 5, place Jules~Janssen, 92195 Meudon, France.}

\abstract
    {Less than~$3\%$ of the known exoplanets were directly imaged for two main reasons. They are angularly very close to their parent star, which is several magnitudes brighter. Direct imaging of exoplanets thus requires a dedicated instrumentation with large telescopes and accurate wavefront control devices for high-angular resolution and coronagraphs for attenuating the stellar light. Coronagraphs are usually chromatic and they cannot perform high-contrast imaging over a wide spectral bandwidth. That chromaticity will be critical for future instruments.}
    {Enlarging the coronagraph spectral range is a challenge for future exoplanet imaging instruments on both space-based and ground-based telescopes.}
    {We propose the multi-stage four-quadrant phase mask that associates several monochromatic four-quadrant phase mask coronagraphs in series. Monochromatic device performance has already been demonstrated and the manufacturing procedures are well-under control since their development for previous instruments on VLT and JWST. The multi-stage implementation simplicity is thus appealing.}
    {We present the instrument principle and we describe the laboratory performance for large spectral bandwidths and for both pupil shapes for space-~(off-axis telescope) and ground-based~(E-ELT) telescopes.}
    {The multi-stage four-quadrant phase mask reduces the stellar flux over a wide spectral range and it is a very good candidate to be associated with a spectrometer for future exoplanet imaging instruments in ground- and space-based observatories.}
    \keywords{Instrumentation: high angular resolution --- Techniques: high angular resolution -- Method: laboratory --- Method: numerical}
    
    \date{ }
    
    \titlerunning{Laboratory demonstration of MFQPM}
    \authorrunning{R.~Galicher et al.}
    
    \maketitle
    
    \section{Introduction}
    Orbital parameters as well as atmosphere physical characteristics of exoplanets orbiting in the outer part of their stellar systems will put strong constraints on planetary formation models. The sole method for studying these objects is direct imaging, which measures both astrometric and spectro-polarimetric parameters. This method is also a direct way to study planet-disk or planet-planet interactions in multiple systems. Up to now only a dozen exoplanet candidates have been imaged in favorable conditions: bright planets orbiting far from their star~(several~AU). When trying to detect fainter planets with that technique, high contrast and small angular separation between the star and its planet become critical points. Large telescopes are thus required to reach the angular resolution, and coronagraphs or interferometers  are needed to reduce the stellar flux. Adaptive optics and speckle calibrations are also necessary to deal with the speckle noise induced by the Earth's atmosphere and optical aberrations. In a few months, the SPHERE~\citep[VLT,][]{beuzit08}, GPI~\citep[Gemini,][]{macintosh08} and HiCIAO~\citep[Subaru,][]{tamura06} instruments will combine coronagraphs and extreme adaptive optics to study  planets~$\sim\!\!\!\!10^{6}$ fainter than their star and measure their spectro-polarimetric characteristics. The main limitation will be the uncorrected atmospheric turbulence by adaptive optics and slowly drifting optical aberrations. For the following generation of high-contrast imaging instruments like~EPICS for~E-ELT~\citep{kasper08} and the~TMT planet finder~\citep{macintosh06}, more accurate adaptive optics and speckle calibrations are planned and coronagraphs with higher efficiency are needed. The objective is to associate coronagraphs and speckle calibrators to reach high contrasts~\citep[$10^6$-$10^7$ stellar attenuations and speckle calibration up to a factor of~$10^2$-$10^3$,][]{verinaud08} over a large spectral band~($\sim\!\!20\%$). Several methods are suggested to build achromatic coronagraphs, especially those that use phase-shift masks~\citep{soummer03,mawet05,mawet06,carlotti09,mawet11}. To achromatize the four-quadrant phase mask~\citep[FQPM,][]{Rouan00}, our team proposed to use several monochromatic devices in series and create the multi-stage four-quadrant phase mask~\citep[MFQPM,][]{baudoz08}. {Because} the monochromatic~FQPM technology is well under control since the developments made for~VLT~\citep{boccaletti04} and~JWST~\citep{boccaletti05}, an~MFQPM is easy to build. We dedicate the present paper to this achromatic coronagraph. We recall the technique principle in~\S\ref{sec : principle} and present a laboratory performance for visible wide spectral bandwidths~($20\%$ and~$30\%$) in~\S\ref{sec : experiment} as well as pupil shapes for space-~(off-axis telescope) and ground-based~(E-ELT) telescopes.

    \section{Principle}
    \label{sec : principle}
    After describing the~FQPM chromaticity limitation in~\S\ref{subsec : fqpm}, we present the multi-stage~FQPM coronagraph principle in~\S\ref{subsec : mfqpm}.
    
    \subsection{Monochromatic FQPM}
    \label{subsec : fqpm}
    We do not present a complete~FQPM study in this section. The device is already well described in~\citet{Rouan00}, \citet{riaud01,riaud03}, and \citet{boccaletti04}. We only remark on properties that are used to design the~MFQPM.
    
\begin{figure}[!ht]
	\centering
      \includegraphics{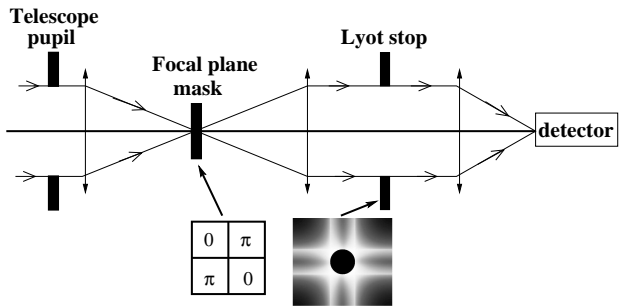}
      \caption[]{\it Single~FQPM coronagraph schematics. The image shows the intensity distribution in the pupil plane just before the Lyot-stop.}
      \label{fig : f1}
    \end{figure}

\begin{figure}[!ht]
    	\centering
	  \includegraphics{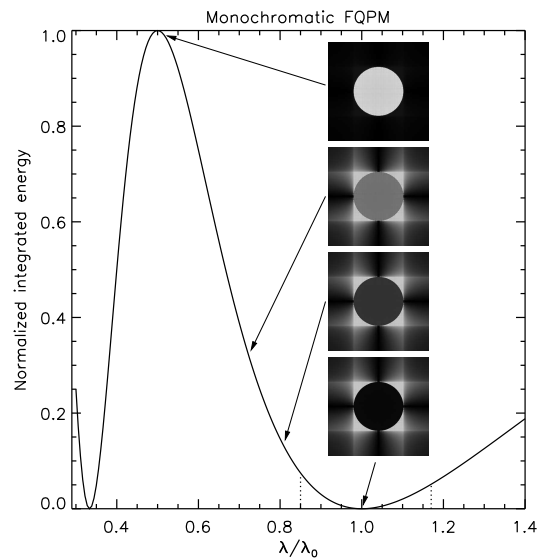}
      \caption[]{\it Integrated energy over the geometrical pupil after a monochromatic~FQPM against wavelength. Intensity distributions just before the Lyot-stop are shown for four wavelengths~(same gray scale). The curve is normalized to the non-coronagraphic case.}
      \label{fig : f2}
    \end{figure}

    The~FQPM is a phase mask that induces a~$\pi$ phase shift on two quadrants of a diagonal and no phase shift in the two other quadrants. In a single~FQPM coronagraph~(Fig.\,\ref{fig : f1}), the on-axis source beam is centered on the phase mask in a focal plane. The resulting diffraction pattern in the following pupil plane is drawn in Fig.\,\ref{fig : f1}. No light goes through the geometrical pupil, so that the on-axis source light, called starlight below, is entirely stopped by the Lyot-stop filtering diaphragm.
    The coronagraph performance mainly depends on the~FQPM quality: the~$\pi$ phase shift accuracy and the transition thickness between the four quadrants. Phase shifts are usually produced by material steps. This solution makes the mask easy to build but has the disadvantage that the mask is monochromatic because the phase shift is exactly~$\pi$ for a unique wavelength, which is called the optimized wavelength. Assuming no wavefront errors, infinitely thin transitions between the quadrants, a full entrance pupil, and no Lyot-stop undersizing, we plot in~Fig.\,\ref{fig : f2} the integrated energy over the geometrical pupil plane as a function of wavelength. We normalize the curve to the integrated energy over the entrance pupil~(non-coronagraphic case).
        The corresponding intensity distributions in the pupil plane before the Lyot-stop are also shown for four wavelengths~(same gray scales). At the optimized wavelength~$\lambda_0$, the entire stellar light is diffracted outside the geometrical pupil and stopped by the Lyot-stop: no startlight goes through the instrument. For other wavelengths~$\lambda$, the phase shift is~$\pi\,\lambda_0/\lambda$ so that a part of the starlight is not diffracted outside the geometrical pupil and the stellar extinction is not perfect~(see pupil images). But this unstopped stellar light is uniformly distributed inside the geometrical pupil in the Lyot-stop plane~(appendix~\ref{app : pupcoro}), which is an interesting feature, as explained in~\S\ref{subsec : mfqpm}. The worst case happens for~$\lambda=\lambda_0/2$, because the phase shift is then~$2\,\pi$~(ie.~$0$) and the entire starlight energy is inside the geometrical pupil after the~FQPM~(no attenuation).

    We define the average transmission~$\tau$ as the ratio of the stellar energy that goes through the Lyot-stop in the spectral bandwidth of interest for the coronagraphic case to the same quantity for the non-coronagraphic case. It represents the average residual stellar energy after the coronagraph. The smaller this number, the more effective the coronagraph. Using a single monochromatic~FQPM,~$\tau$ is the integrated area under the curve in~Fig.\,\ref{fig : f2} bounded by the extremal bandwidth wavelengths and divided by the bandwidth. If the bandwidth is~$32\%$~(vertical dotted lines), $\tau$ is about~$2.10^{-2}$. If it is~$20\%$, $\tau$ is~$8.10^{-3}$. These stellar attenuations are not sufficient when exoplanet detections require at least~$\tau<10^{-4}$. Several solutions have been proposed to build achromatic~FQPM~\citep[e.g., ][]{bloemhof05,mawet06,carlotti09}. The multi-stage four-quadrant phase mask~\citep{baudoz08} that we study in this paper is a part of these solutions.
    
    \subsection{Multi-stage FQPM}
    \label{subsec : mfqpm}
\begin{figure}[!ht]
 	\centering
     \includegraphics{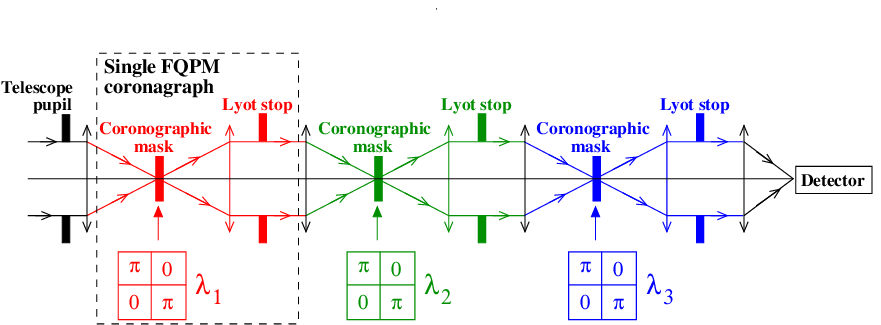}
      \caption[]{\it MFQPM coronagraph schematics. Three monochromatic~FQPM are used.}
      \label{fig : f3}
    \end{figure}

\begin{figure}[!ht]
   	\centering
  \includegraphics{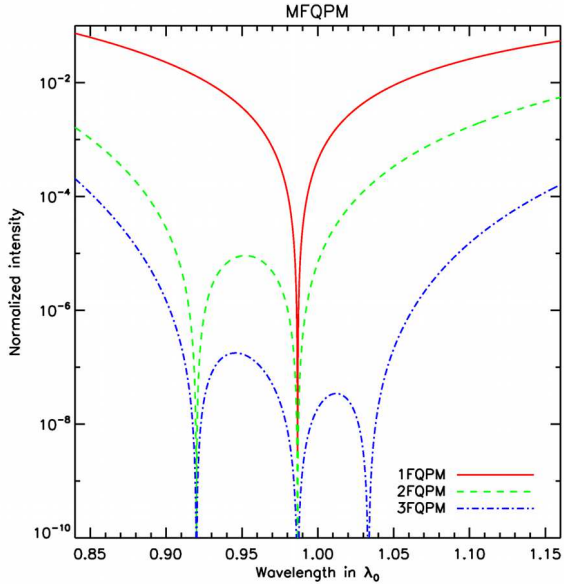}
      \caption[]{\it Residual energy spectra after one~(red full line), two~(green dashed line), and three~(blue dashed-dotted line) monochromatic~FQPM. The curves are normalized to the non-coronagraphic case.}
      \label{fig : f4}
    \end{figure}

    Let us consider a single~FQPM that perfectly cancels the light at its optimized wavelength~$\lambda_1$. At another wavelength~(i.e.~$\lambda\ne\lambda_1$), the residual starlight is uniformly distributed inside the Lyot-stop geometrical pupil. Thus, a second monochromatic~FQPM can be used in series to reduce the residual starlight around a second optimized wavelength~$\lambda_2\ne\lambda_1$. After the second stage, the unaffected light by both the first and the second~FQPM is still uniformly distributed inside the geometrical pupil. A third stage optimized at~$\lambda_3$ can be used, and so on. We call this monochromatic coronagraph combination multi-stage four-quadrant phase mask coronagraph~(Fig.\,\ref{fig : f3}). 
   Note that the combination works because the~FQPM unaffected light pupil after the mask is equal to the pupil before the mask up to a factor of proportionality. The following focal mask can then deal with it~(appendix~\ref{app : pupcoro}). Calling~$\lambda_0$ the central wavelength of the spectral band, we plot the integrated energy in the third Lyot-stop pupil for a MFQPM composed by three single~FQPM respectively optimized for~$\lambda_1=0.987\,\lambda_0$, $\lambda_2=0.920\,\lambda_0$ and $\lambda_3=1.033\,\lambda_1$~(blue dashed-dotted line in~Fig.\,\ref{fig : f4}). We choose these optimized wavelengths to match the experiment case presented in section~\ref{sec : experiment}. We overplot the curves for the first coronagraph used alone~(red full line) and associated with the second mask~(two~FQPM in series, green dashed line). All curves are normalized to the non-coronagraphic case. In both~Fig.\,\ref{fig : f2} and~\ref{fig : f4} we assume no wavefront errors, infinitly thin~FQPM transitions, a full pupil and no Lyot-stop undersizing. The vertical scale is now logarithmic, whereas it is linear in~Fig.\,\ref{fig : f2}. Each monochromatic mask is more effective at its optimized wavelength than at the others. Each mask enhances the contrast over the whole spectral range, and three monochromatic~FQPM in series~(blue curve) with a full pupil and no aberrations give an average transmission~$\tau$ as low as~$\sim\!\!10^{-6}$~($\sim\!\!2.10^{-5}$) in a~$20\%$~($32\%$) bandwidth.
   
   \section{Laboratory performance}
   \label{sec : experiment}
This section describes laboratory experiments. We first present the actual~MFQPM prototype~(\S\ref{subsec : proto}). We then give the performance using a full pupil (\S\ref{subsec : fullpup}) and an obstructed pupil (\S\ref{subsec : eelt}).

\subsection{MFQPM prototype}
\label{subsec : proto}
\begin{figure}[!ht]
	\centering
     \includegraphics{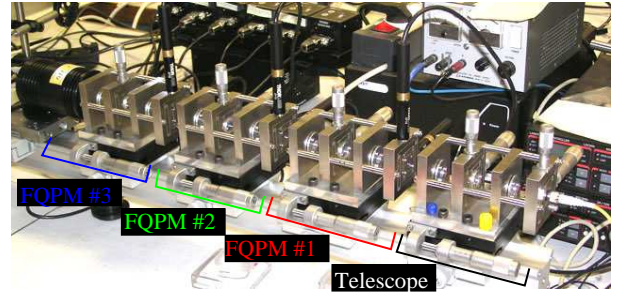}
	      \caption[]{\it MFQPM prototype picture. A first stage simulates the telescope and three monochromatic~FQPM are in series.}
      \label{fig : f5}
    \end{figure}

A first prototype was used to demonstrate the~MFQPM principle using a full circular pupil~($D$-diameter) a few years ago~\citep{baudoz08}. The main result was the detection of a~$7.10^{-9}$ laboratory companion~(i.e., not numerically added a posteriori) at~$4.5\,\lambda_0/D$ in a~$20\%$ visible band after a post-processing differential imaging to enhance the raw~MFQPM contrast~($7.10^{-7}$ for a signal-to-noise ratio of~$3$). But that prototype worked with a slow beam~(f$/300$) and was sizable~($2.5$\,m in length). We thus designed a compact prototype with a~$40$ optical f-number, which is more compatible with a real instrument~(Fig.~\ref{fig : f5}).
Each~FQPM stage is composed of one monochromatic focal mask, one collimating lens, one Lyot-stop and one imaging lens. A translation mechanism moves each stage along the optical axis with a micrometric positioning. The full instrument length is less than~$45$\,cm~($<\!\!15$\,cm per stage). All lenses are achromatic doublets. Coatings reduce the reflected light to less than~$0.1\%$ in energy from~$625$\,nm to $825$\,nm at each lens surface. The entering beam optical aperture is f$/40$ at each stage and full or obstructed pupil shapes can be used. The three monochromatic~FQPM are optimized at~$\lambda_1=740$\,nm, $\lambda_2=690$\,nm, and $\lambda_3=775$\,nm~(numbering is according to their position from the light source to the detector). Each~FQPM transversal position is controlled by two motorized positioning stages. The source creates an ultrabroadband supercontinuum radiation in the visible domain. A beamsplitter is set between the third stage and the detector. One beam converges on the detector~(CCD) for imaging. The second beam feeds an optical fiber linked to a spectrometer. The fiber diameter is~$\lambda_0/D$~($\lambda_0=750$\,nm) and its mount enables accurate movements so that spectra at different positions in the focal plane can be recorded. We use that laboratory setup for both experiments in  the full pupil~(\S\ref{subsec : fullpup}), and the obstructed pupil~(\S\ref{subsec : eelt}).

   \subsection{Full circular pupil}
\label{subsec : fullpup}
We first test the~MFQPM prototype with a full circular entrance pupil. A slight Lyot-stop diameter undersizing is required to efficiently block the central source light. We use a~$90\%$ filtering at each stage. The instrument throughput is then~$73\%$. Table~\ref{tab : tab1} gives the diaphragm sizes.
\renewcommand{\arraystretch}{1.2}
\begin{table}[!ht]
\centering
\begin{tabular}{c|c||c|c}
Entrance Pupil&$780\,\mu$m &Lyot $2$&$630\,\mu$m\\
\hline
Lyot $1$&$700\,\mu$m&Lyot $3$&$570\,\mu$m\\
\end{tabular}
\caption[]{\it Diaphragm diameters for the full pupil experiment.}
\label{tab : tab1}
\end{table}

\begin{figure}[!ht]
	\centering
    \includegraphics{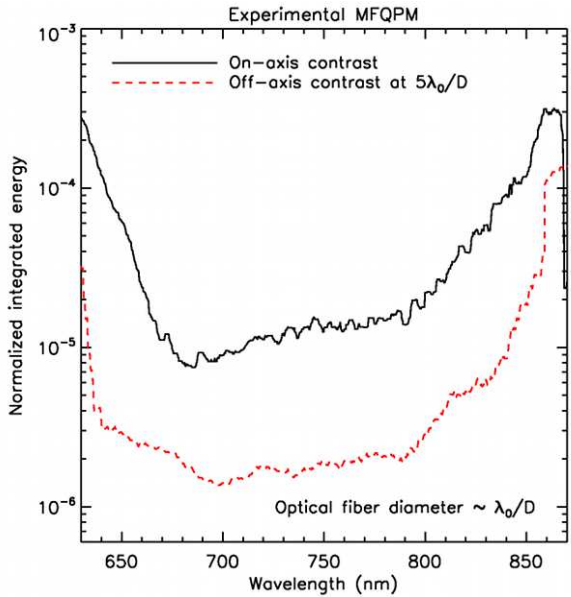}
      \caption[]{\it Experimental residual energy spectra integrated by the spectrometer fiber on the optical axis~(black full line) and at~$5\,\lambda_0/D$~(red dashed line) in the focal plane of a~MFQPM with a full pupil. The curves are normalized to the non-coronagraphic case.}
      \label{fig : f6}
    \end{figure}

We focus the first study on the spectral MFQPM performance. We record the energy entering the spectrometer fiber when the~MFQPM is aligned. We plot that quantity as a function of wavelength when the fiber is centered on the optical axis~(black full line in~Fig.\,\ref{fig : f6}). We normalize the curve to the energy entering the fiber in the non-coronagraphic case~(focal masks are pushed off-center). 
The coronagraph gives an average transmission between~$7.10^{-6}$ and~$4.10^{-5}$ at each wavelength over a~$20\%$ bandwidth~($660-800$\,nm). That demonstrates the~MFQPM capacity to achromatically reduce the starlight over a wide spectral range. However, the performance~(see Tab\,\ref{tab : tab2}) is higher than the expected ones from numerical predictions~(Fig.\,\ref{fig : f4}, factor  up to~$10$).
\begin{table}[!ht]
\centering
\begin{tabular}{c|c|c}
\multirow{2}*{$\Delta\lambda$}&\multicolumn{2}{c}{Average Transmission~$\tau$}\\
\cline{2-3}
&On-axis&At $5\,\lambda_0/D$\\
\hline
$20\%$&$1.4\,10^{-5}$&$2.0\,10^{-6}$\\
\hline
$32\%$&$3.6\,10^{-5}$&$6.1\,10^{-6}$\\
\end{tabular}
\caption[]{\it Experimental performance of actual~MFQPM in full pupil.}
\label{tab : tab2}
\end{table}
We investigated the possible actual prototype limitations. We first measured from microscopic images that the transition thickness between the focal mask quadrants is about~$1\,\mu$m. We found from numerical simulations developed for previous studies~\citep[JWST,][]{baudoz06b} that this thickness set a contrast limit between~$10^{-6}$ and~$10^{-5}$. Because our bench is not located in a cleanroom environment, the dust has to be taken into account as well to estimate the individual~FQPM performance, especially given the small pupil size. A dust density estimation over the pupil area was made using microscopic images at the different surfaces of one doublet lens. We constructed one transmission map per~FQPM stage by multiplying all images. These amplitude maps were added in the pupil planes assuming all recorded dusts are utterly dark. We then used our numerical simulations of the complete MFQPM to predict a residual diffracted energy level up to $10^{-5}$ into the fiber.
Finally, the mid-frequencies errors introduced by the lenses were estimated by measuring surface errors on one lens with a white-light interferometric intrument. Considering the substrate optical index, they reach about~$1.45$\,nm~rms for a single achromatic doublet lens. Even accounting for all lenses, these small phase errors limit the contrast to~$10^{-7}$. It seems then that dusts and transitions are the two main current prototype limitations. A careful lens fabrication process and a protective cover over the instrument should be sufficient to decrease the dust effect. We will also work to obtain thinner~FQPM transitions.

In the second full pupil~MFQPM experiment, the spectrometer fiber is set at~$5\,\lambda_0/D$ from the optical axis and at~$45\degree$ from the~FQPM transitions~(the three mask transitions are parallel).  The residual energy spectrum is plotted as a red dashed line in~Fig.\,\ref{fig : f6} with the same normalization as the other curve. As expected, the level is lower than for the on-axis case. This level is limited however by speckles that can be detected in the images and that probably come from the limitations described above. A~$2\,10^{-6}$ planet spectrum could be detected in the raw coronagraphic image with a signal-to-noise ratio~(SNR) of~$3$ over a~$20\%$ bandwidth. That raw performance will be ameliorated using the new prototype with less dust and transition effects (under fabrication). A speckle calibration will also enable a contrast gain of up to several magnitudes.

   \subsection{Obstructed pupil}
    \label{subsec : eelt}
\begin{figure}[!ht]
	\centering
    \includegraphics{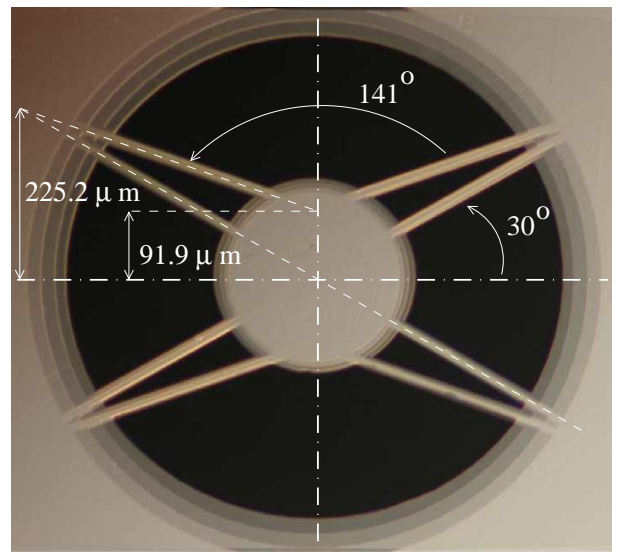}
      \caption[]{\it Superimposition of microscope images of the E-ELT pupil and the three associated~$95\%$ Lyot-stops used in experiment.}
      \label{fig : f7}
    \end{figure}

In this section, we report the~MFQPM experimental performance when used behind an obstructed telescope. Because the study is realized in the framework of the future European~ELT~(E-ELT) planet finder, ~EPICS~\citep{kasper08}, the experimental telescope pupil is one of those that are planned for the~E-ELT, though our setup does not simulate primary mirror segmentation. We found from numerical simulations made for E-ELT studies that gaps between segments diffract light farther than~$30\,\lambda/D$ and that their impact is negligible for the coronagraphic performance. The experimental pupil microscope image is presented in~Fig.\,\ref{fig : f7}. We superimpose in the same figure the three~$95\%$ Lyot-stop pictures.
    All diaphragms are composed of a central obstruction and eight spiders. Both the vertical and horizontal axis are symmetric axes. The spider positions are indicated in the figure, and Table\,\ref{tab : tab3} describes diameters and spider thickness. The instrument throughput is~$86\%$~(Lyot-stop undersizing).
\begin{table}[!ht]
\centering
\begin{tabular}{c|c|c|c}
\multirow{2}*{Diaphragm}&Internal&External&Spider\\
&Diameter ($\mu$m)&Diameter ($\mu$m) &Thickness ($\mu$m)\\
\hline
Telescope&$231$&$780$&$9.3$\\
\hline
Lyot $1$&$243$&$741$&$11$\\
\hline
Lyot $2$&$257$&$702$&$13$\\
\hline
Lyot $3$&$270$&$669$&$15$\\
\end{tabular}
\caption[]{\it E-ELT diaphragm parameters.}
\label{tab : tab3}
\end{table}

\begin{figure}[!ht]
	\centering
     \includegraphics{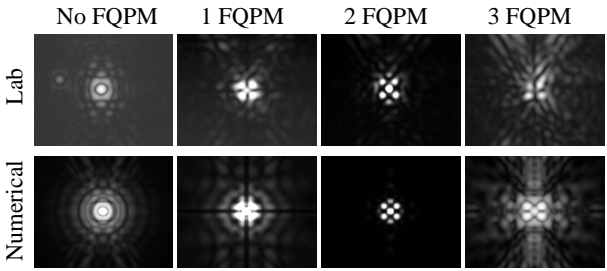}
  \caption[]{\it First row: laboratory images with~$0$, $1$, $2$, and $3$~FQPM in series for an~E-ELT pupil. Second row: numerical simulations.}
      \label{fig : f8}
    \end{figure}

\begin{figure}[!ht]
	\centering
\includegraphics{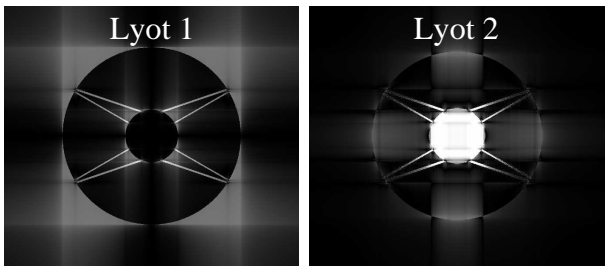}
      \caption[]{\it Energy distribution before the first~(left) and the second~(right) Lyot-stops in corresponding pupil planes for an obstructed pupil.}
      \label{fig : f9}
    \end{figure}

\begin{figure}[!ht]
	\centering
\includegraphics{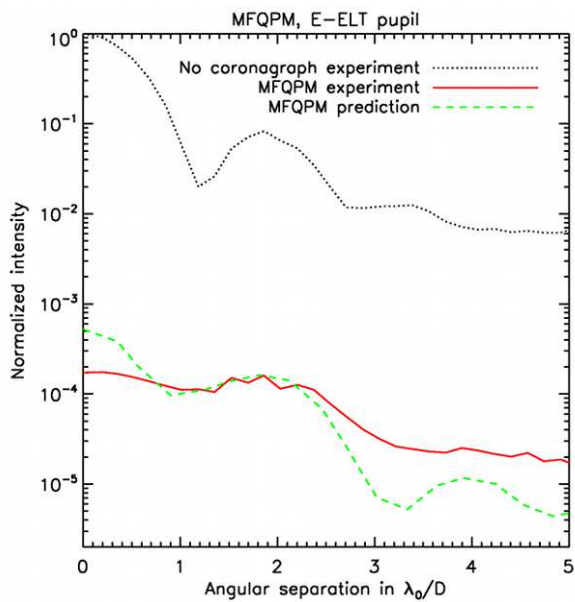}
      \caption[]{\it Experimental intensity against angular separation after no coronagraph~(black dotted line) or  MFQPM~(red full line) in visible~($630$ to~$870$\,nm.). The numerical prediction is shown in a green dashed line. The curves are normalized to the non-coronagraphic~PSF maximum. Entrance pupil is the~E-ELT pupil.}
      \label{fig : f10}
    \end{figure}

Experimental broadband~($630$ to~$870$\,nm) images recorded after aligning no coronagraph, the first~FQPM, the two first~FQPM, and the three~FQPM~(complete~MFQPM) are presented in the first row in~Fig\,\ref{fig : f8}. They are similar to the numerically simulated images drawn in the second row. 
With no coronagraph, the image is the~E-ELT point spread function~(PSF) exhibiting a central core and spider diffraction spikes. After a single~FQPM, the pattern has a Maltese cross shape. The light whose wavelength is around the  phase mask optimized wavelength~$\lambda_1$ is affected by the coronagraph and is spread in the cross-braces. The four peak pattern is the~FQPM characteristic image when the pupil has a central obstruction. For wavelengths different from~$\lambda_1$, the light is almost unaffected by the~FQPM and it is diffracted in a pattern very similar to the~E-ELT~PSF: bright central part of the Maltese cross. When a second~FQPM is added, it is not easy to distinguish which part of the image caused by light affected or unaffected by the phase masks. It is interesting however to study the energy distribution before the first and the second Lyot-stops in the corresponding pupil planes~(Fig.~\ref{fig : f9}).
	Let us call~$D_o$ the entrance obstruction diameter. In the first Lyot-stop plane, the pupil obstruction diffraction pattern created by the first~FQPM is similar to the~$D_o$-diameter circular pupil diffraction pattern~(Babinet's principle). The image on the left in~Fig.\,\ref{fig : f9} shows the light that goes through the first Lyot-stop around the central obstruction. This explains why the single~FQPM performance is degraded compared to the case with no obstruction. In the focal plane, the leak creates the four peaks of the Maltese cross~(second column in~Fig.~\ref{fig : f8}). Consider now that a second~FQPM is used in series. It diffracts the light that is diffracted by the first mask, so that in the second Lyot-stop plane, the obstruction-diffraction pattern is the obstruction itself~(right in~Fig.\,\ref{fig : f9}). All light that leaks through the first Lyot-stop is then blocked by the second Lyot diaphragm obstruction. The same property is used in the multi-stage vortex coronagraph~\citep{mawet11}. More precisely, a part of the obstruction diffracted light still goes through the second Lyot-stop because the first Lyot-stop filters the  obstruction diffraction pattern~(finite external diameter), and the first and second focal masks are not optimized for the same wavelengths. The central obstruction bad effect is very well attenuated however, and the~MFQPM performance is very attractive even with a~$30\%$ obstruction in diameter as for the~E-ELT pupil we used~(see below). Finally, like the image produced by a single~FQPM, the image recorded after the third~FQPM~(complete~MFQPM, Fig.~\ref{fig : f8}) reveals a four-peak pattern close to the center because a small part of the light is affected by the obstruction of the second Lyot-stop~(odd number of~FQPM). However because the flux was attenuated at all the wavelengths by the instrument, no bright peak remains in the center unlike in the single~FQPM case. We also notice that the image quality beyond than~$2\,\lambda_0/D$ is limited by speckles at a level specified in~Fig.\,\ref{fig : f10}~(experimental curve above predictions made with no speckle noise). The curves represent the experimental image intensity profiles recorded without coronagraph~(black dotted line) and after the~MFQPM prototype~(red full line) against the angular separation to the central source. A numerical simulation prediction is overplotted in a green dashed line~(no phase errors and infinitely thin transitions). All curves are normalized to the~E-ELT~PSF maximum intensity~(no-coronagraphic case) and correspond to the intensity integrated over a~$32\%$ band~($630$ to~$870$\,nm).
The average transmission~$\tau$ in the~$32\%$ band is~$\sim\!\!4.10^{-4}$, and from the contrast curve, we find that a~$5.9\,10^{-5}$~($5.9\,10^{-4}$)  companion can be detected as close as~$5\,\lambda_0/D$~($2\,\lambda_0/D$) with a~SNR of~$3$.

We then recorded spectra under the same experimental conditions. 
    \begin{figure}[!ht]
	\centering
\includegraphics{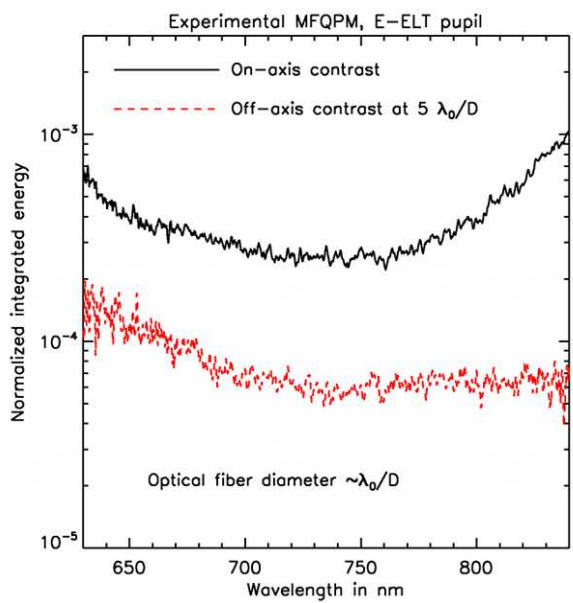}
      \caption[]{\it Experimental residual energy spectral evolution integrated by the spectrometer fiber on the optical axis~(black full line) and at~$5\,\lambda_0/D$~(red dashed line) in MFQPM focal plane with~E-ELT pupil. The curves are normalized to the non-coronagraphic case.}
      \label{fig : f11}
    \end{figure}
    Curves that are normalized to the non-coronagraphic spectrum are drawn in~Fig.~\ref{fig : f11}. On the optical axis~(black full line), the~MFQPM attenuates the source and leaves a residual energy as low as~$3.2\,10^{-4}$ from~$660$ to~$800$\,nm~($20\%$ bandwidth, see~Tab.~\ref{tab : tab4}).
    \begin{table}[!ht]
\centering
\begin{tabular}{c|c|c}
\multirow{2}*{$\Delta\lambda$}&\multicolumn{2}{c}{Average Transmission~$\tau$}\\
\cline{2-3}
&On-axis&At $5\,\lambda_0/D$\\
\hline
$20\%$&$3.2\,10^{-4}$&$6.4\,10^{-5}$\\
\hline
$32\%$&$3.8\,10^{-4}$&$7.1\,10^{-5}$\\
\end{tabular}
\caption[]{\it Current MFQPM experimental performance with~E-ELT pupil.}
\label{tab : tab4}
\end{table}
At~$5\,\lambda_0/D$~(red dashed line), the instrument can record a~$2\,10^{-4}$ companion spectrum over the~$30\%$ band~($600$-$870$\,nm) with a~SNR of~$3$. Again, that performance is extracted from the raw coronagraphic data and, as the image quality is limited by speckle noise further than a few~$\lambda_0/D$~(Fig~\ref{fig : f8}), better contrasts can be reached by applying speckle calibrations.

       \section{Conclusions}
       We presented the current multi-stage four-quadrant phase mask~\citep[MFQPM,][]{baudoz08} laboratory performance. The instrument uses several monochromatic four-quadrant phase masks~\citep[FQPM][]{Rouan00} in series. It is easy to build because the monochromatic FQPM fabrication is very well under control since the developments made for~JWST and~VLT instruments.
              
       In the laboratory, we demonstrated that the~MFQPM strongly attenuates the central source over a very wide spectral range~($>20\%$) with a high throughput~($86\%$) even as close as~$2\,\lambda_0/D$ from the star. Using a~f$/40$ off-axis telescope~(full pupil), the actual prototype can record a~$2.10^{-6}$ planet spectrum at~$5\,\lambda_0/D$ from its star over a~$20\%$ bandwidth~($660-800$\,nm) with a~SNR of~$3$.
       
       We also proved that the~MFQPM is not very sensitive to ground telescope central obstructions and spiders unlike a lot of coronagraphs. Using an~E-ELT pupil and a~f$/40$ optical aperture, the actual prototype can measure a~$2.\,10^{-4}$ planet spectrum at~$5\,\lambda_0/D$ over a~$20\%$ bandwidth with a~SNR of~$3$.
        
        Finally, it is very important to recall that all these results of the raw~MFQPM performance were made under the current laboratory conditions~(no cleanroom). As explained in~\S\ref{subsec : fullpup} and~\S\ref{subsec : eelt}, actual limitations are dusts and~FQPM transition thickness and phase aberrations  at a lower level. To reach higher contrasts~(up to~$10^{10}$ at a few~$\lambda_0/D$), a new prototype is under fabrication and we will use it in a cleanroom. We will also associate the~MFQPM with a~self-coherent camera~\citep{baudoz06,galicher10} to both actively~(deformable mirror) and a posteriori reduce the speckle noise.
       
       \appendix
       \section{Light distribution in a FQPM Lyot-stop}
       \label{app : pupcoro}
The focal plane mask coronagraph effect in the output pupil plane at the wavelength~$\lambda$ can be described by
\begin{equation}
\psi_\lambda'(\xi,\nu)=F\left[F^{-1}\left[\psi_\lambda\right]\,M_\lambda(x,y)\right]\,L(\xi,\nu),
\label{eq : puplyot}
\end{equation}
with $F$ and $F^{-1}$ the Fourier transform and its inverse, $\psi_\lambda$ and $\psi_\lambda'$ the pupil plane complex amplitudes before and after the focal mask~$M_\lambda$, $L$ the Lyot-stop function, $(\xi,\nu)$ the pupil coordinates, and $(x,y)$ the focal plane coordinates. The perfect~FQPM mask function optimized for the wavelength~$\lambda_1$ but working at an other wavelength~$\lambda$ can be described by
\begin{equation}
M_\lambda(x,y)= \exp{\left[i\,\phi(\lambda)\,(1-sgn(x)\, sgn(y))/2\right]},
\label{eq : fqpmmask}
\end{equation}
where~$sgn(x)$ is the sign of~$x$ and~$\phi(\lambda)=\pi\,\lambda_1/\lambda$ is the phase shift induced between the quadrants. Developing~Eq.\,\ref{eq : fqpmmask}, we find
\begin{eqnarray}
M_\lambda(x,y)= \exp{\left[i\,\phi(\lambda)/2\right]}\,\big(\cos{(\phi(\lambda)/2)}&\label{eq : fqpmmask2}\\
-i\,sgn(x)\,sgn(y)&\sin{(\phi(\lambda)/2)}\big)\nonumber,
\end{eqnarray}
We deduce the field in the output pupil for the wavelength~$\lambda$ from~Eqs.\,\ref{eq : puplyot} and~\ref{eq : fqpmmask2}
\begin{eqnarray}
\psi_\lambda'(\xi,\nu)=\exp{\left[i\,\phi(\lambda)/2\right]}\,\Big[\cos{(\phi(\lambda)/2)}\,\psi_\lambda(\xi,\nu)\,L(\xi,\nu)\\
-i\,\sin{(\phi(\lambda)/2)}\,F\left[F^{-1}[\psi_\lambda]\,sgn(x)\,sgn(y)\right]\,L(\xi,\nu)\Big],
\end{eqnarray}
We recognize two terms multiplied by the phasor~$\exp{\left[i\,\phi(\lambda)/2\right]}$. The first term describes a pupil equal to the entrance pupil up to a factor of proportionality  and a possible undersizing induced by the Lyot-stop. It corresponds to the light that is not affected by the focal mask. It goes to zero as the wavelength tends to~$\lambda_1$~($\phi$ tends to~$\pi$). The second term represents the pupil affected by a perfect~FQPM weighted by a constant. Assuming perfect optics, an infinite field and an unobstructed pupil, the term is nulled~\citep{abe03}. In that case, we find
\begin{equation}
\psi_\lambda'(\xi,\nu)=\exp{\left[i\,\phi(\lambda)/2\right]}\,\cos{(\phi(\lambda)/2)}\,\psi_\lambda(\xi,\nu)\,L(\xi,\nu),
\end{equation}
	The pupil distribution at~$\lambda$ after a~FQPM optimized at~$\lambda_1$ is the same as in the entrance pupil but weighted by~$\cos{[\pi\,\lambda_1/(2\,\lambda)]}$. A second~FQPM coronagraph can then be applied to reduce this residual light. Considering~$n$ coronagraphs in series optimized for~$\lambda_{1\le i\le n}$ and no Lyot-stop undersizing, the ratio~$\tau$ of the integrated energy with and without~MFQPM over the bandwidth~$[\lambda_{min},\lambda_{max}]$ is
\begin{equation}
\tau =\int_{\lambda_{min}}^{\lambda_{max}} \prod_{i=1}^{i=n} \cos^2{\left(\frac{\pi}{2}\,\frac{\lambda_i}{\lambda}\right)}\,\mathrm{d}\lambda.
\end{equation}

\end{document}